\documentclass[twocolumn]{article}
\usepackage{latexsym}
\usepackage{amsmath}
\usepackage{amssymb}
\usepackage{graphicx}
\usepackage{cite}
\usepackage{longtable}
\date{}

\title{Numerical solution of the Boltzmann equation for the shock
wave in a gas mixture}
\author{A. A. Raines\\ 
Faculty of Mathematics and Mechanics, St. Petersburg State University\\
Universitetskii pr. 28, Staryi Peterhof, St.-Petersburg 198504, Russia\\
{\small E-mail: raines@mail.ru}}
\date{}

\begin{document}
\maketitle

\begin{abstract}
We study the structure of a shock wave for a two-, three- and
four-component gas mixture on the basis of numerical
solution of the Boltzmann equation for the model of hard sphere
molecules. For the evaluation of collision integrals we
use the Conservative Projection Method developed by F.G. Tscheremissine
which we extended to gas mixtures in cylindrical coordinates.
The transition from the upstream to downstream uniform state is presented by macroscopic values and
distribution functions. The obtained results were compared with
numerical and experimental results of other authors.
\end{abstract}

\textbf{List of symbols}\vspace*{2mm}\\
\begin{tabular}{ll}
  $m_i$                      & molecular mass of $i$-th component\\
  $\vec{p}_i$                & molecular momentum\\
                             & of $i$-th component\\
  $\vec{x}$                  & vector of configuration space\\
  $f_i(\vec{p}_i,\vec{x},t)$ & distribution function\\
                             & of $i$-th component\\[1mm]
  $\vec{g}_i = \frac{\displaystyle\vec{p}_j}{\displaystyle m_2} - \frac{\displaystyle\vec{p}_i}{\displaystyle m_1}$ & initial relative velocity\\[1mm]
  $(p_{x,i},p_{r,i})$        & coordinates of momentum\\
                             & in cylindrical coordinates\\
  $M$                        & Mach number\\
  $R$                        & gas constant\\
\end{tabular}

\begin{tabular}{ll}
  $k$                        & Boltzmann constant\\
  $p_\gamma$                 & nodes of momentum grids\\
  $\Omega$                   & domain in momentum space\\
  $V$                        & volume of domain $\Omega$\\
  $N_0$                      & number of momentum nodes\\
  $N_\nu$                    & number of integration nodes\\
  $d_i$                      & diameter of molecules of $i$-th component\\
  $n_i$                      & number density of $i$-th component\\
  $u_i$                      & flow velocity of $i$-th component\\
  $T_i$                      & temperature of $i$-th component\\
  $T_{xx,i}$                 & parallel temperature of $i$-th component\\
  $T_{rr,i}$                 & transversal temperature of $i$-th component\\
  $n$                        & number density of the mixture\\
  $\rho$                     & mass density of the mixture\\
  $u$                        & flow velocity of the mixture\\
  $T$                        & temperature of the mixture\\
  $l$                        & mean free path of molecules\\
  $\tau$                     & mean collision time\\
  $\chi_i$                   & concentration  of $i$-th component
\end{tabular}

\section{Introduction}

The shock wave structure for a binary gas mixture is an important
problem in kinetic theory. It has been investigated experimentally
and theoretically by using moment methods, direct simulation
Monte-Carlo method (DSMC), fluid dynamics methods, numerical
analysis based on kinetic models, conservative splitting method,
finite-difference analysis of the Boltzmann equation (see a review
in \cite{RainesEu}).

The first numerical solution of the Boltzmann equation for a single gas was obtained in \cite{nord} and then in \cite{Tcherem70}. Later, the problem was solved by discrete-ordinate methods for Boltzmann equation with different techniques of evaluation of the collision integral: polynomial approximation of distribution function in velocity space \cite{Oh93,Oh94}, application of polynomial correction for fulfilling conservation laws \cite{Arist80}, conservative projection method \cite{Cherem97,Tcherem98,Tcherem00,Tcherem06}.

The method of \cite{Oh94} was extended to binary gas mixtures \cite{Kosuge}, the method of polynomial correction \cite{Arist80} was applied to binary gas mixtures in \cite{Maus,Raines91}, the method of \cite{Tcherem98,Tcherem00} was extended at first to binary gas mixtures \cite{RainesEu,Raines02} and then to three- and four-component mixtures \cite{Raines09}. Later computations for 3-component mixtures were repeated by the same method in \cite{Jos}.

In this paper the shock wave structure for two- three- and four-component gas mixtures is solved by an extension of Conservative Projection Method \cite{Cherem97,Tcherem98,Tcherem00,Tcherem06} applied to the complete kinetic Boltzmann equation. The method is based on a special projection techniques for evaluation of the collision integrals. The computed collision integral is conservative for density, momentum and energy. It is equal to zero when the solution has a form of Maxwellian distribution function. The integration grid for evaluation of the collision integral is given in \cite{korob}. The differential part of the Boltzmann equation is approximated by conservative finite-difference scheme of the second order \cite{Boris}. In this scheme the transport of mass, momentum and energy between the nodes of the configuration space is realized in a conservative way.

Here we have studied the behavior of densities, flow velocities, parallel, transversal and total temperatures for the mixture and its components for various Mach numbers, various masses and concentration ratios. The results were compared with the results of \cite{Kosuge} and experiments \cite{Harnett} with a good agreement between them.

\section{Description of the method}

The system of Boltzmann kinetic equations for a mixture of monatomic  gases containing $K$ components is usually written in the form
\begin{equation}
\frac{\partial F_i}{\partial t} +
\vec{\xi}_i\frac{\partial F_i}{\partial \vec{x}} = I_i,\quad i = 1,\dots,K.
 \label{1}
\end{equation}
The collision integrals have the form
\begin{equation}
 I_i =
  \sum_j\int\limits_{\mathbb{R}^{\,3}}\int\limits_0^{2\pi}\int\limits_0^{b_m}\left(F^\prime_i F^\prime_j - F_iF_j\right) g\, b\, d\,b\, d\varepsilon
\, d\vec{\xi}_j .
  \label{2}
\end{equation}
Here we have used the notation: $F_i = F_i(\vec{\xi}_i,\vec{x},t)$, $F^\prime_i = F_i(\vec{\xi}^\prime_i,\vec{x},t)$, $\vec{\xi}_i, \vec{\xi}_j$ and $\vec{\xi}^\prime_i, \vec{\xi}^\prime_j$ are velocity vectors before and after collisions, respectively,  $g = |\vec{\xi}_j - \vec{\xi}_i|$, $b_m$ is the maximum interaction distance, $b$ and $\varepsilon$ are impact parameters of a binary collision. To extend the conservative method of evaluating collision integrals \cite{Tcherem06} to gas mixtures, it is sufficient to transform equation (\ref{1}) from the velocity variables to momentum variables:
\begin{equation*}
    (\vec{\xi}_i,\vec{x},t)\mapsto (\vec{p}_i,\vec{x},t),\quad F_i = F_i(\vec{\xi}_i,\vec{x},t)\mapsto f_i(\vec{p}_i,\vec{x},t).
\end{equation*}
From the normalization condition $\int F_i\, d\,\vec{\xi}_i\! = \int f_i\, d\,\vec{p}_i = n_i$, one obtains $F_i(\vec{\xi}_i,\vec{x},t)=f_i(\vec{p}_i,\vec{x},t)m_i^3$. The system of Boltzmann equations in the momentum space take the form
\begin{equation}\label{3}
 \frac{\partial f_i}{\partial t} + \frac{\vec{p}_i}{m_i}\,\frac{\partial f_i}{\partial \vec{x}} = I_i .
\end{equation}
Collision integrals (\ref{2}) become
\begin{equation}
 I_i =
  \sum_j\int\limits_{\mathbb{R}^{\,3}}\int\limits_0^{2\pi}\int\limits_0^{b_m}\left(f^\prime_i f^\prime_j - f_if_j\right) g\, b\, d\,b\, d\varepsilon
\, d\vec{p}_j ,
  \label{4}
\end{equation}
where
\[f_i = f_i(\vec{p}_i,\vec{x},t),\; f^\prime_i = f_i(\vec{p}^{\,\,\prime}_i,\vec{x},t),\; g = \left|\frac{\vec{p}_j}{m_2} - \frac{\vec{p}_i}{m_1}\right|.\]
The following properties should be conserved in a discrete form of the collision integral:
\begin{gather*}
\int\limits_{\mathbb{R}^{\,3}}I_i(\vec{p}_i)\, \Psi(\vec{p}_i)\, d\,\vec{p}_i = 0,\;\textrm{where}\quad \Psi(\vec{p}_i) = \left(1,\vec{p}_i,\frac{\vec{p}_i^2}{m_i} \right) \\
I_i\left[f_{i,M}\right] = 0,\;\textrm{where}\\
f_{i,M} = n_i\left(\frac{1}{2\pi k T m_i}\right)^{3/2}\exp{\left(-\frac{(\vec{p}_i-\vec{p}_{i0})^2}{m_i 2kT}\right)}.
\end{gather*}
System (\ref{3}) with collision integrals (\ref{4}) is solved either on a uniform 3-dimensional grid $S_0$ with $N_0$ points $p_\gamma$ in Cartesian momentum space $\Omega$ of volume $V$ or on a uniform 2-dimensional grid in cylindrical coordinate system due to the cylindrical symmetry of the problem. For brevity, values of collision integrals and distribution functions at the grid nodes are denoted by $I_{i,\gamma}$ and $f_{\alpha,\gamma}\ (\alpha=i,j)$, respectively. The use of a constant step in the coordinate space is needed for conservation of the total momentum in the projection method.

System (\ref{3}) of $K$ equations is transformed to the system of $N_0K$ equations:
\begin{gather}
 \frac{\partial f_{i,\gamma}}{\partial t} + \frac{\vec{p}_{i,\gamma}}{m_i}\,\frac{\partial f_{i,\gamma}}{\partial \vec{x}} = I_{i,\gamma},\notag \\ \qquad\qquad \qquad  i = 1,2,\dots,K,\quad \gamma = 1,2,\dots,N_0.
 \label{9}
\end{gather}

Evaluation of collision integrals is performed with the use of 8-dimensional uniform integration grid $S_\nu=(\vec{p}_{i,\nu},\vec{p}_{j,\nu},b_\nu,\varepsilon_\nu)$ in the domain $\Omega\times\Omega\times [0,2\pi]\times [0,b_m]$ with $N_\nu$ nodes (the cross $\times$ denotes direct product) in such a way that momenta $\vec{p}_{i,\nu}$ and $\vec{p}_{j,\nu}$ coincide with momentum grid nodes while all the variables $b_\nu,\varepsilon_\nu$, for which the post-collision momentum $\vec{p}^\prime_{i,\nu}$ or $\vec{p}^\prime_{j,\nu}$ falls outside of $\Omega$, are excluded. The collision integral for $n$-th component at node $\gamma$ can be written in the form
\begin{equation}\label{5}
    I_{n,\gamma} = \frac{1}{4}\sum\limits_{i,j} \int\limits_{\Omega\times\Omega}\int\limits_0^{2\pi}\int\limits_0^{b_m} \Phi_{n,\gamma}(f^\prime_i f^\prime_j - f_i f_j) g b d b d\varepsilon d\vec{p}_i d\vec{p}_j
\end{equation}
with $\Phi_{n,\gamma}$ being the following combination of Dirac $\delta$-functions and Kroneker symbols $\delta_{n,l}$ ($\delta_{n,l}=1$ if $n=l$ and $\delta_{n,l}=0$ if $n\neq l$):
\begin{gather}
\Phi_{n,\gamma} = \delta_{n,i}\delta(\vec{p}_i - \vec{p}_{i,\gamma}) + \delta_{n,j}\delta(\vec{p}_j - \vec{p}_{j,\gamma})\notag \\
\mbox{}\qquad - \delta_{n,i}\delta(\vec{p}^{\,\,\prime}_i - \vec{p}_{i,\gamma}) - \delta_{n,j}\delta(\vec{p}^{\,\,\prime}_j - \vec{p}_{j,\gamma}).
\label{6}
\end{gather}

The conservative projection method for evaluation of (\ref{5}) imply replacing the two last $\delta$-functions in (\ref{6}) by their decompositions with a splitting coefficient $r_\nu\leqslant1$ which has to be defined from the energy conservation law. For each contribution to the integral sum this decomposition has the form (omitting subscript $\nu$):
\begin{gather}
\delta(\vec{p}^{\,\,\prime}_i - \vec{p}_{i,\gamma}) = (1-r)\delta(\vec{p}_{i,\lambda} - \vec{p}_{i,\gamma}) + r\delta(\vec{p}_{i,\lambda+s} - \vec{p}_{i,\gamma})
\notag \\
\delta(\vec{p}^{\,\,\prime}_j - \vec{p}_{j,\gamma}) = (1-r)\delta(\vec{p}_{j,\mu} - \vec{p}_{j,\gamma}) + r\delta(\vec{p}_{j,\mu-s} - \vec{p}_{j,\gamma})
\label{7}
\end{gather}
In (\ref{7}) the grid nodes $\vec{p}_{i,\lambda}$ and $\vec{p}_{j,\mu}$ are the closest ones to the post-collision vectors $\vec{p}^{\,\,\prime}_i$ and $\vec{p}^{\,\,\prime} _j$, respectively, whereas $\vec{p}_{i,\lambda+s}$ and $\vec{p}_{j,\mu-s}$ are some complementary nearly located grid nodes. Hence the contributions to the collision integral in two near-grid points are replaced by the weighted contributions in two pairs of the closest nodes. A necessary condition to make this decomposition conservative is the fulfillment of the momentum conservation law. Using a uniform grid in $\Omega$ for the kinetic equation in the momentum space ensures the fulfillment of this condition.

Coefficient $r$ (subscript $\nu$ is skipped) is defined from the energy conservation law
\begin{gather}
 E_0 = (1-r)E_1 + rE_2,\quad \textrm{where}\quad E_0 = \frac{\vec{p}_i^{\,2}}{2m_i} + \frac{\vec{p}_j^{\,2}}{2m_j},\notag \\
 E_1 =  \frac{\vec{p}_{i,\lambda}^{\,2}}{2m_i} + \frac{\vec{p}_{j,\mu}^{\,2}}{2m_j},\quad E_2 = \frac{\vec{p}_{i,\lambda+s}^{\,2}}{2m_i} + \frac{\vec{p}_{j,\mu-s}^{\,2}}{2m_j}.
 \label{8}
\end{gather}

\subsection{\textbf{Details of calculations in cylindrical coordinates}}

For computing the shock wave structure it is convenient to pass to cylindrical coordinate system $(p_x,p_r,\varphi)$ in momentum space in which the distribution function $f_i(\vec{p}_i,\vec{x},t)$ is replaced by the function $f_i(\vec{p}_{x,i},\vec{p}_{r,i},\vec{x},t)$ independent of $\varphi$ due to the assumed cylindrical symmetry of the problem. In domain $\Omega$ of volume $V$ in momentum space we introduce a uniform two-dimensional grid with $N_0$ points $(p_{x,\gamma},p_{r,\gamma})$. Equation (\ref{9}) becomes
\begin{gather*}
 \frac{\partial f_{i,\gamma}}{\partial t} + \frac{p_{x,i,\gamma}}{m_i}\,\frac{\partial f_{i,\gamma}}{\partial x} = I^c_{i,\gamma},\notag \\ \qquad\qquad \qquad  i = 1,2,\dots,K,\quad \gamma = 1,2,\dots,N_0.
\end{gather*}
For the evaluation of collision integrals we use the grid
\[S^c_\nu = \{(p_{x,i,\nu},p_{r,i,\nu}),(p_{x,j,\nu},p_{r,j,\nu}),\varphi_{\nu},\vartheta_{\nu},b_\nu,\varepsilon_\nu\}\]
with $N_\nu$ nodes in such a way that $(p_{x,i,\nu},p_{r,i,\nu})$ and $(p_{x,j,\nu},p_{r,j,\nu})$ coincide with the momentum grid nodes, while the angles are distributed uniformly in corresponding intervals. The integral (\ref{5}) must be written in cylindrical coordinates:
\begin{gather*}
    I^c_{n,\gamma} = \frac{1}{4}\sum\limits_i\sum\limits_j \int\limits_{\Omega\times\Omega}\int\limits_0^{2\pi}\int\limits_0^{2\pi}\int\limits_0^{2\pi}\int\limits_0^{b_m} \Phi^c_{n,\gamma}(f^\prime_i f^\prime_j - f_i f_j) g\\ \qquad\qquad\qquad\qquad\quad \times p_{r,i} p_{r,j} d\, p_{r,i} d\, p_{r,j} b\, d\, b\, d\,\varphi d\,\vartheta\, d\,\varepsilon .
\end{gather*}
In formulas (\ref{6}) and (\ref{7}) we make the following replacements
\begin{gather*}
  \delta(\vec{p}_i - \vec{p}_{i,\gamma}) = \delta(p_{x,i} - p_{x,i,\gamma}) \delta(p_{r,i} - p_{r,i,\gamma}),\\
    \delta(\vec{p}_j - \vec{p}_{j,\gamma}) = \delta(p_{x,j} - p_{x,j,\gamma}) \delta(p_{r,j} - p_{r,j,\gamma})
\end{gather*}
and
\begin{gather*}
  \delta(\vec{p}^{\,\,\prime}_i - \vec{p}_{i,\gamma}) = (1 - q)\delta(p_{x,i,\lambda} - p_{x,i,\gamma}) \delta(p_{r,i,\lambda} - p_{r,i,\gamma})\\
  \mbox{} + q\delta(p_{x,i,\lambda+s} - p_{x,i,\gamma}) \delta(p_{r,i,\lambda+s} - p_{r,i,\gamma}),\\
    \delta(\vec{p}^{\,\,\prime}_j - \vec{p}_{j,\gamma}) = (1 - q)\delta(p_{x,j,\mu} - p_{x,j,\gamma}) \delta(p_{r,j,\mu} - p_{r,j,\gamma})\\
  \mbox{} + q\delta(p_{x,i,\mu-s} - p_{x,j,\gamma}) \delta(p_{r,j,\mu+\tilde{s}} - p_{r,j,\gamma}),
\end{gather*}
respectively. Here the nodes nearest to $\vec{p}^{\,\,\prime}_i$ and $\vec{p}^{\,\,\prime}_j$ are $(p_{x,i,\lambda},p_{r,i,\lambda})$ and $(p_{x,j,\mu},p_{r,j,\mu})$, respectively. The two sets of four nodes surrounding the points $\vec{p}^{\,\,\prime}_i$ and $\vec{p}^{\,\,\prime}_j$ can be denoted by $(p_{x,i,\lambda+s},p_{r,i,\lambda+s})$ and $(p_{x,j,\mu-s},p_{r,j,\mu+\tilde{s}})$, where $\vec{s}$ and $\tilde{\vec{s}}$ are vectors of displacement along the grid. In Cartesian coordinates from the well-known relation between vectors of momentum before and after the collision we have $\vec{s} = - \tilde{\vec{s}}$. However, in cylindrical coordinates we have only one equality for the first $x$-coordinate.

The decomposition coefficient $q_\nu$ ($\nu$ has been skipped in the formulas above and below) can be determined from the energy conservation law:
\[  E_0 = E_1(1-q) + qE_2,\] where
\begin{gather*}
  E_0 = \frac{p_{x,i}^2+p_{r,i}^2}{2m_i} + \frac{p_{x,j}^2+p_{r,j}^2}{2m_j} ,\\
  E_1 = \frac{p_{x,i,\lambda}^2+p_{r,i,\lambda}^2}{2m_i} + \frac{p_{x,j,\mu}^2+p_{r,j,\mu}^2}{2m_j} ,\\
  E_2 = \frac{p_{x,i,\lambda+s}^2+p_{r,i,\lambda+s}^2}{2m_i} + \frac{p_{x,j,\mu-s}^2+p_{r,j,\mu+\tilde{s}}^2}{2m_j}.
\end{gather*}

We have demonstrated in some detail how a transformation of variables from velocity space to momentum space (in Cartesian and cylindrical coordinates) in the system of Boltzmann equations makes it possible to build the conservative projection method for evaluation of collision integrals.

After calculating the collision integral, the system of discrete ordinate equations (\ref{9}) is solved by the standard procedure of applying the splitting method until the stabilization of the solution.

The obtained distribution functions define the following gas dynamics parameters for components: the number densities $n_i$, flow velocities $u_i$, temperatures $T_i$, parallel temperatures $T_{xx,i}$, and transversal (radial) temperatures $T_{rr,i}$. For the whole gas, one obtains the molecular number density $n$, density $\rho$, flow velocity $u$, and temperature $T$. The listed macroscopic variables are defined as sums of moments of the distribution functions. For a gas mixture in cylindrical coordinates one has
\begin{gather*}
  n_i = \sum\limits_\gamma p_{r,i,\gamma} f_{i,\gamma},\quad u_i = \frac{1}{n_im_i}\sum\limits_\gamma p_{r,i,\gamma}p_{x,i,\gamma}f_{i,\gamma},\\
  T_{xx,i} = \frac{1}{kn_im_i}\sum\limits_\gamma p_{r,i,\gamma} (p_{x,i,\gamma} - u_{x,i}m_i)^2 f_{i,\gamma},\\
  T_{rr,i} = \frac{1}{2kn_im_i}\sum\limits_\gamma p_{r,i,\gamma}^3 f_{i,\gamma},\\
  T_i = \frac{1}{3} (2T_{rr,i} + T_{xx,i}),\quad n = \sum\limits_i n_i,\\
  \rho = \sum\limits_i m_i n_i,\quad u = \frac{1}{\rho} \sum\limits_i m_i n_i u_i,\\
  T = \frac{1}{kn} \sum\limits_i \left[ku_iT_i + m_in_i (u_i - u)^2/3\right] .
\end{gather*}

\section{The problem of a shock wave structure}

We consider a plain shock wave traveling in $x$ direction with Mach number $M$. We denote the parameters: numerical densities, flow velocity and temperature before the shock wave as $n_i^{(1)}$, $n^{(1)}$, $u^{(1)}$, $T^{(1)}$, respectively, and those behind the shock wave as $n_i^{(2)}$, $n^{(2)}$, $u^{(2)}$, $T^{(2)}$. Parameters on both sides of the shock wave are related by the Rankine-Hugoniot conditions for monatomic gas
\begin{gather*}
  \frac{n_i^{(2)}}{n_i^{(1)}} = \frac{u^{(1)}}{u^{(2)}} = \frac{4M^2}{M^2 + 3}\,,\\
  \frac{T^{(2)}}{T^{(1)}} = \frac{\big(5M^2 - 1\big)\big(M^2 + 3\big)}{16M^2}\,,
\end{gather*}
$u^{(1)} = Mc^{(1)}$, $c^{(1)} = \sqrt{5 R^{(1)}T^{(1)}/3}$, $R^{(1)} = k/(m_1\chi_1^{(1)} + m_2\chi_2^{(1)})$, $\chi_i^{(1)} = n_i^{(1)}/n^{(1)}$, $n^{(1)} = n_1^{(1)} + n_2^{(1)}$, where $k$ is the Boltzmann constant, $m_1$, $m_2$ are the masses of the first and second component, respectively, $\chi_1^{(1)}$ and $\chi_2^{(1)}$ are concentrations of the first and second components before the shock wave. Here $M$ is Mach number before the shock wave.

The problem is solved in the coordinate system attached to shock wave.
The steady shock wave structure is obtained as the evolution of initial discontinuity of gas parameters posed at $x=0$. Boundary conditions are imposed at sufficiently large distances from the discontinuity at $x=-L_1$, $x=L_2$, where the gas can be considered as being in thermodynamic equilibrium with the corresponding Maxwellian distribution functions:
\begin{gather*}
  f_i(t=0,x<0,p_{x,i},p_{r,i}) = f_i(t,x=-L_1,p_{x,i},p_{r,i})\\
  = \frac{n_i^{(1)}}{(2\pi T^{(1)}m_i)^{3/2}}\exp{\left(-\frac{(p_{x,i}-u^{(1)}m_i)^2 + p_{r,i}^2}{2kT^{(1)}m_i}\right)},\\
    f_i(t=0,x>0,p_{x,i},p_{r,i}) = f_i(t,x=L_2,p_{x,i},p_{r,i})\\
  = \frac{n_i^{(2)}}{(2\pi T^{(2)}m_i)^{3/2}}\exp{\left(-\frac{(p_{x,i}-u^{(2)}m_i)^2 + p_{r,i}^2}{2kT^{(2)}m_i}\right)}.
\end{gather*}
The function $f_i(t>0,-L_1<x<L_2,p_{x,i},p_{r,i})$ is searched by solving the Boltzmann equation. After the solution is found at the velocity grid $S_0 = (p_{x,\gamma},p_{r,\gamma})$, gas parameters are computed as the sums given at the end of the previous section.

\section{Presentation of macroscopic gas parameters}

For the presentation of macroscopic gas parameters we use two forms: normalized form and reduced form. Parameters presented in the first form are normalized by their values ahead of the shock wave: $n_i/n^{(1)}$, $T_i/T^{(1)}$ and so on. In the second form computed results are presented by the following reduced parameters (with the asterisks further removed):
\begin{gather*}
n_i^\ast = (n_i-n_i^{(1)})/(n_i^{(2)} - n_i^{(1)}),\\
T_i^\ast = (T_i-T_i^{(1)})/(T_i^{(2)} - T_i^{(1)}),\\
u_i^\ast = (u_i-u_i^{(1)})/(u_i^{(1)} - u_i^{(2)})
\end{gather*}
and similarly for the parameters of the mixture $n, u ,T$.

Characteristic parameters are the free path and velocity:
\begin{gather*}
  l^{(1)} = \left(\sqrt{2}\pi n^{(1)} d_1^2\right)^{-1},\quad c_0 = \sqrt{2kT^{(1)}/m_1},\\
  \tau = l^{(1)}/c_0, \quad p_{\,0} = m_1c_0,
\end{gather*}
where $l^{(1)}$ is the mean free path of molecules of the first component in an equilibrium state at rest with the number density $n^{(1)}$.

\section{Numerical results}

We consider the shock wave structure in a mixture of monatomic gases assuming the hard-sphere model of molecules. The shock wave structure in the mixture is defined by a number of dimensionless parameters: Mach number $M$, concentrations $\chi_i^{(1)} = (n_i^{(1)}/n^{(1)})$ of gas components before the shock wave, molecular mass ratio $m_i/m_1$, and ratio of molecular diameters $d_i/d_1$; $m_1$ and $d_1$ are parameters of the first component, $n^{(1)}$ is the density of the mixture before the shock wave.

\begin{figure}
\begin{center}
\includegraphics[width=0.33\textwidth,height=!,angle=-90,origin=c]{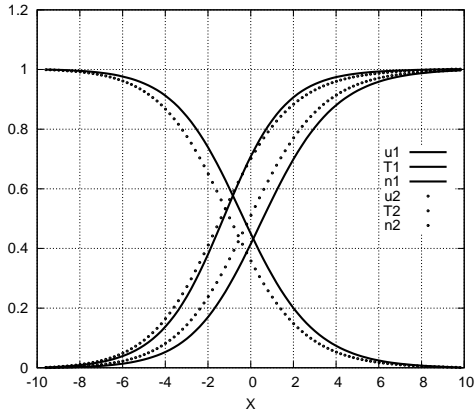}
\vspace*{-6mm}
\caption{\footnotesize Shock wave structure for two component mixture \newline\hspace*{9mm} at $M = 1.5$, $m_2/m_1 = 0.5$,
$\chi_2^{(1)} = 0.9$.} \label{fig:1}
\end{center}
\end{figure}

\begin{figure}
\begin{center}
\includegraphics[width=0.33\textwidth,height=!,angle=-90,origin=c]{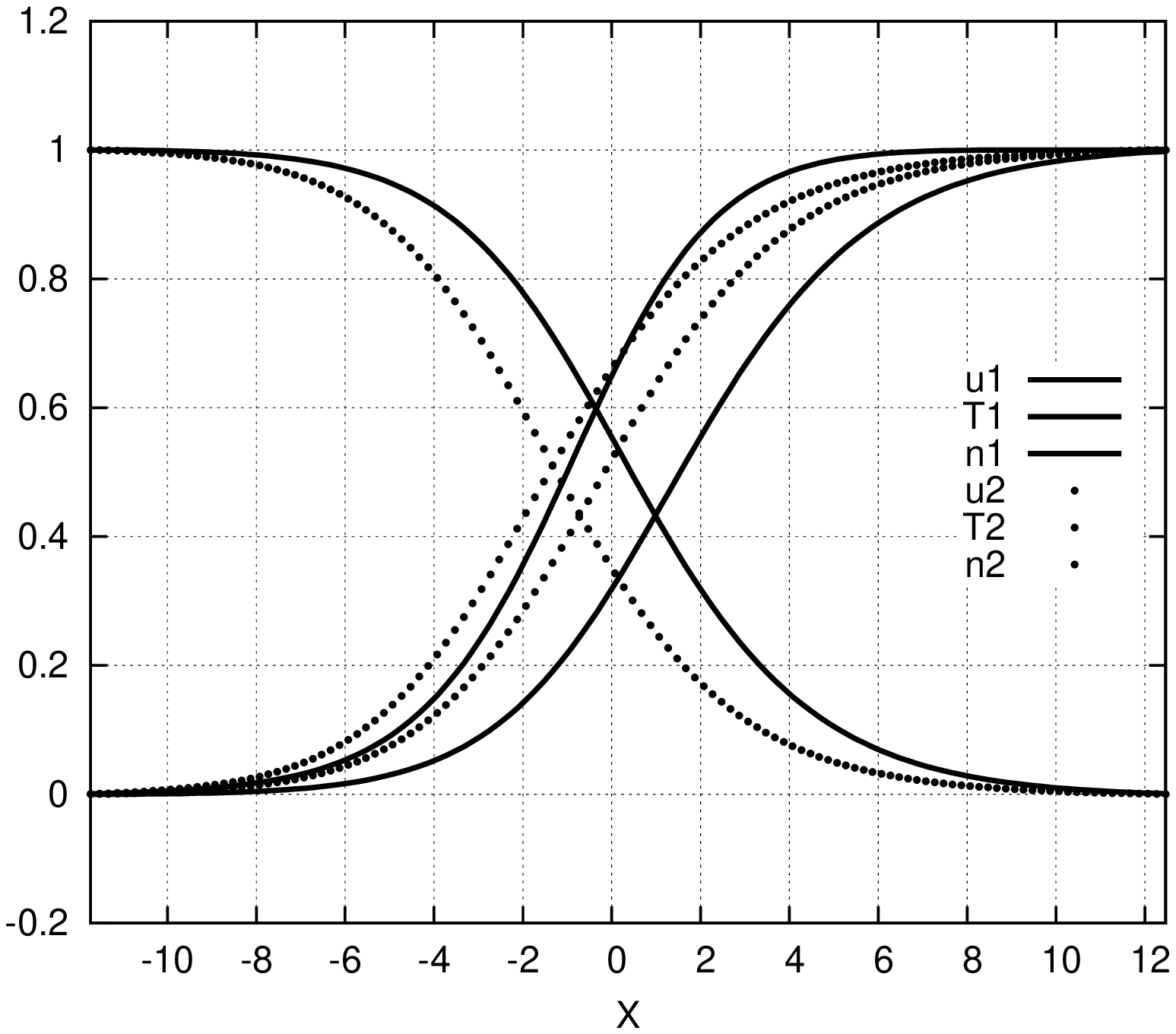}
\vspace*{-6mm}
\caption{\footnotesize Shock wave structure for two component mixture \newline\hspace*{9mm} at $M = 1.5$, $m_2/m_1 = 0.25$,
$\chi_2^{(1)} = 0.9$.} \label{fig:2}
\end{center}
\end{figure}

\begin{figure}
\begin{center}
\includegraphics[width=0.33\textwidth,height=!,angle=-90,origin=c]{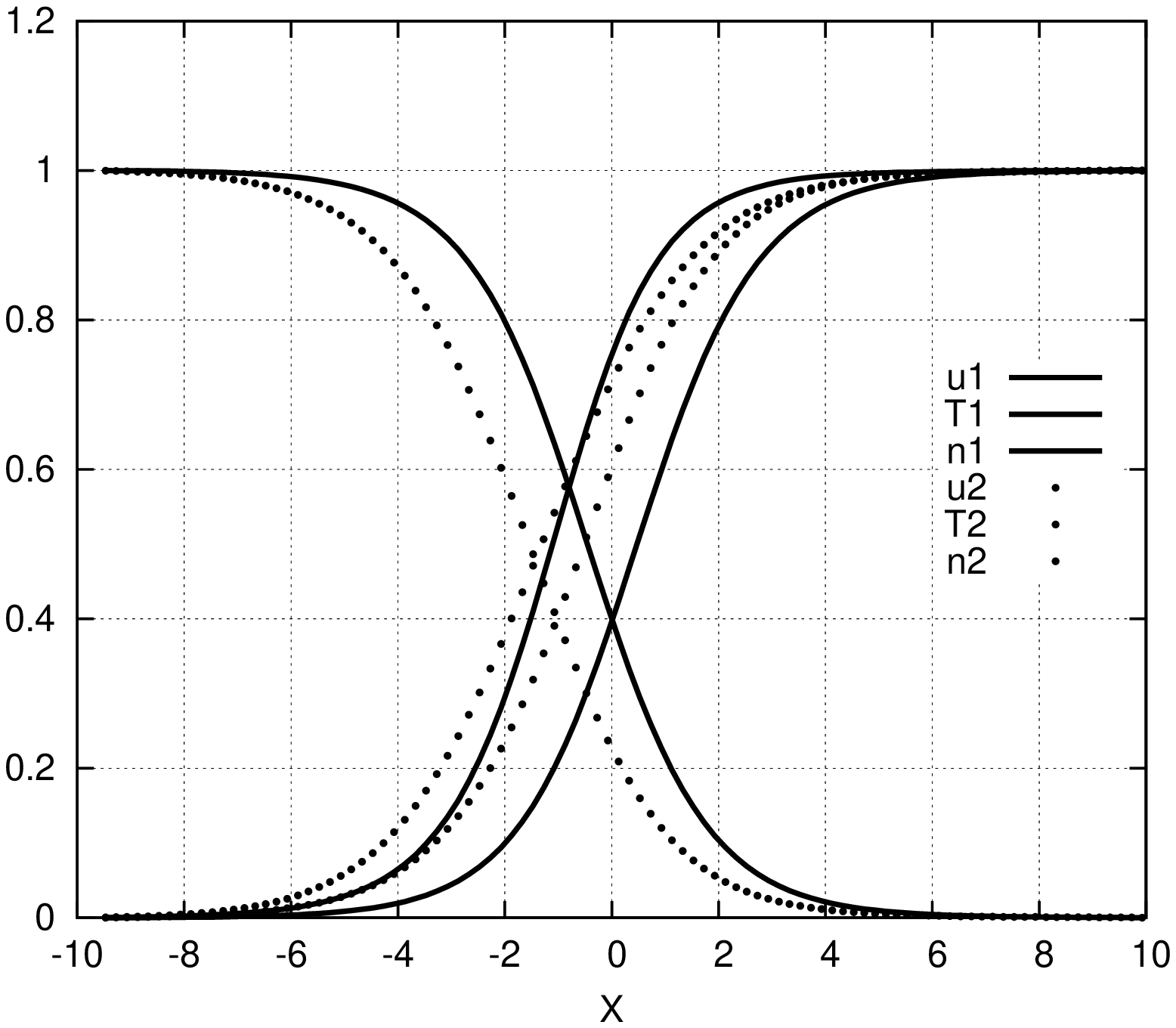}
\vspace*{-6mm}
\caption{\footnotesize Shock wave structure for two component mixture \newline\hspace*{9mm} at $M = 2$, $m_2/m_1 = 0.25$,
$\chi_2^{(1)} = 0.5$.} \label{fig:3}
\end{center}
\end{figure}

\begin{figure}
\begin{center}
\includegraphics[width=0.33\textwidth,height=!,angle=-90,origin=c]{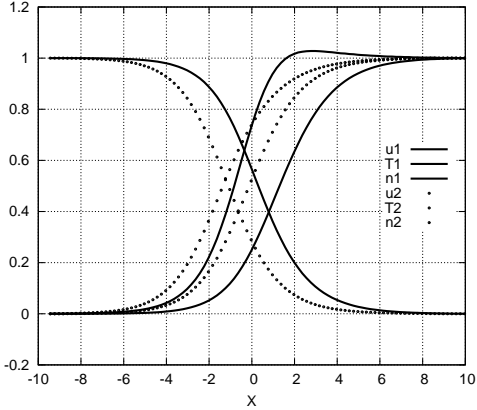}
\vspace*{-6mm}
\caption{\footnotesize Shock wave structure for two component mixture \newline\hspace*{9mm} at $M = 2$, $m_2/m_1 = 0.25$,
$\chi_2^{(1)} = 0.9$.} \label{fig:4}
\end{center}
\end{figure}

\begin{figure}
\begin{center}
\includegraphics[width=0.33\textwidth,height=!,angle=-90,origin=c]{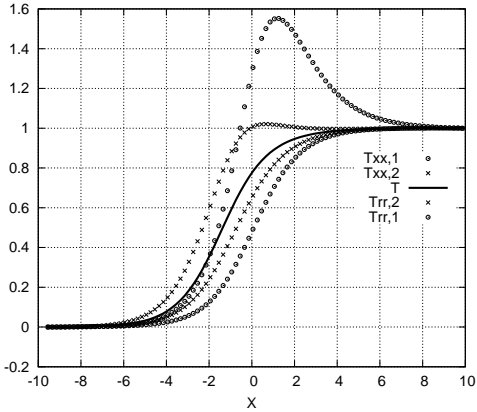}
\vspace*{-6mm}
\caption{\footnotesize Temperature tensors for  $M = 2$, $m_2/m_1 = 0.25$,\newline\hspace*{9mm}
 $\chi_2^{(1)} = 0.95$.}
 \label{fig:5}
 \end{center}
\end{figure}

\begin{figure}
\begin{center}
\includegraphics[width=0.33\textwidth,height=!,angle=-90,origin=c]{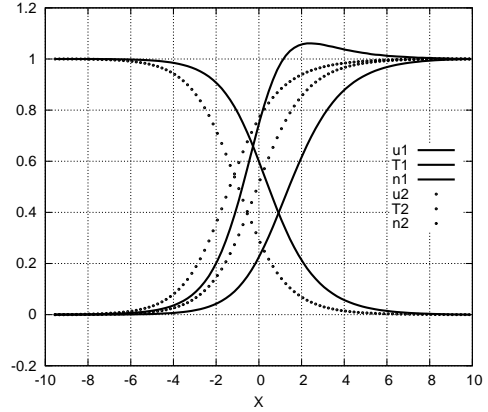}
\vspace*{-6mm}
\caption{\footnotesize Shock wave structure for two component mixture\newline\hspace*{9mm}  at $M = 2$,
$m_2/m_1 = 0.25$, $\chi_2^{(1)} = 0.95$.}
 \label{fig:6}
 \end{center}
\end{figure}

\begin{figure}
\begin{center}
\includegraphics[width=0.33\textwidth,height=!,angle=-90,origin=c]{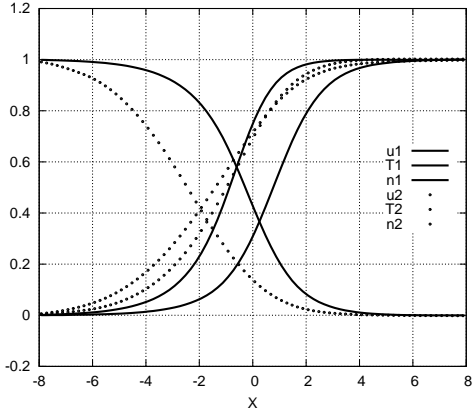}
\vspace*{-6mm}
\caption{\footnotesize Shock wave structure for $M = 3$,
$m_2/m_1 = 0.1$,\newline\hspace*{9mm}  $\chi_2^{(1)} = 0.5$.}
 \label{fig:7}
 \end{center}
\end{figure}

The ratio of diameters plays relatively small role because real molecular diameters are close to each other. The mass ratio and concentrations of components affect relaxation processes inside the shock wave and form its structure.The presentation of macroscopic values is given in the two forms: normalized form and reduced form. We use the same notation for the variables in two forms: $n_i$ and $n$ for number densities, $u_i$ and $u$ for flow velocities, $T_i$ and $T$ for temperatures, $T_{xx,i}$ and $T_{rr,i}$ for parallel and transversal temperatures, respectively. Captions to the figures correspond to curves listed from top left to top right of each figure. In figures \ref{fig:1}--\ref{fig:7} we present results of calculations for a binary mixture with $d_i/d_1 = 1$. A heavy component is considered as the first one.

In figures Fig. \ref{fig:1} and Fig. \ref{fig:2} one can see that the differences between the profiles of densities and temperatures of components increase when the mass ratio decreases. For both cases temperature curves of the heavy component have steeper slopes than that of the light component.
The temperature $T_1$ of the heavy component rises more quickly than the temperature of the light component $T_2$ and exceeds it at some point inside the shock wave. Then $T_1$ either approaches downstream equilibrium temperature monotonously or becomes higher than the downstream temperature and then decreases. Monotonous behavior is seen in Fig. \ref{fig:1}, \ref{fig:2} and \ref{fig:3}, where the influence of concentrations is shown. Fig. \ref{fig:4} shows the non-monotonous behavior of the temperature which becomes apparent at low concentrations of the heavy component and Mach number not too small. This phenomenon had already been discovered by computations in early studies \cite{Bey,bird68} and is known as a temperature overshoot \cite{bird94,bird84}.


Fig. \ref{fig:5} shows the components of temperature tensors for the two constituents of the mixture. The higher hump at the parallel temperature graph of the heavy gas can be explained by inertia of heavy molecules that penetrate more easily into the depth of the shock wave layer when the collisions with light molecules prevail. A contribution of the parallel component with the big hump in $T_1$ yields the overshoot of this temperature. A comparison of curves in Fig. \ref{fig:6} and Fig. \ref{fig:4} shows that the temperature overshoot increases with the rise of concentration of the light gas.

Fig. \ref{fig:7} presents the shock wave structure for the low mass ratio $m_2/m_1 = 0.1$. One can see a big difference of densities and temperatures of components. The temperature profile of the  heavy component is much steeper than that of the light one.

\begin{figure}
\begin{center}
\includegraphics[width=0.33\textwidth,height=!,angle=-90,origin=c]{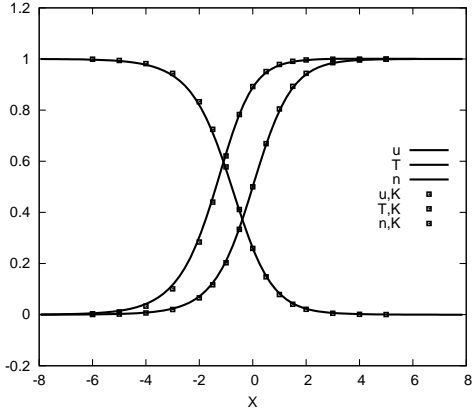}
\vspace*{-6mm}
\caption{\footnotesize Comparison of computations for a binary gas mixture\newline\hspace*{9mm} with \cite{Kosuge} at $M = 3$,
$m_2/m_1 = 0.5$, $\chi_2^{(1)} = 0.9$.}
 \label{fig:8}
 \end{center}
\end{figure}

Fig. \ref{fig:8} presents  a comparison of our results for numerical density, flow velocity and temperature with computations by a different discrete ordinate method in \cite{Kosuge}. Results of \cite{Kosuge} are denoted by squares while our results are shown by solid curves. One can see a good agreement between both of the methods.

 In \cite{Kosuge} the computing time for one iteration step in a parallel computation, using ten CPUs on Fujitsu VPP800 computer, is $142 s$ for $M = 2$ and $99 s$ for $M = 3$. The computer memory for $M = 2$ is $1.7 GB$ and for $M = 3$ is $1.4 GB$. In this work computations were made on a personal computer with processor Pentium 4 with the frequency $2.53\, GHz$ and the memory $512 Mb$. The computing time for one iteration step for $M = 2$ is $14 s$ and for $M = 3$ is $5 s$. The computer memory is $8.5 Mb$. For example, for $Mach = 3, m_2/m_1 = 0.5, d_2/d_1 = 1$ and various concentrations we take the following
values of parameters: $11858$ nodes of the momentum grid with the step $h = 0.1$, $90$ nodes of the $x$-grid with the step $h_x = 0.2$, $198000$ integration nodes and $\Delta t = 0.01$. With this method we can obtain the results on rough grids ($1800$ nodes of the momentum grid with the step $h = 0.26$, 90 nodes of the $x$-grid with the step $h_x = 0.2$, $66000$ integration nodes and $\Delta t = 0.01$) with the computing time for one iteration step $1.8 s$.

The details of the present calculations and analysis of their accuracy can be found in our papers \cite{Raines02}. Calculations for big Mach numbers require large intervals for cylindrical coordinates in momentum space and large intervals in configuration space.

The accuracy of calculations was estimated by comparing
macroscopic quantities for different grids and different numbers
of integration nodes. Let $\sigma(M,S)$ represent either $n$ or
$U$ or $T$ that are obtained using the grids $M$ and $S$, where
$M$ is the momentum grid and $S$ is the grid in configuration
space. We introduce the maximum difference between two results for
two different grids $M, S$ and $M^\prime, S^\prime$ using the
formula
\[D(M^\prime\!,S^\prime\!,M,S)\! = \!\!\!
\max_{\sigma =
n,U,T}\!\left(\max_{x_k}\frac{|\sigma(M^\prime,S^\prime) -
\sigma(M,S)|}{\sigma(M,S)}\right)\] An analogous comparison was
carried out for various numbers of integration nodes and fixed $M$
and $S$. The accuracy of computations is $D = 10^{-2}-10^{-3}$.
Details of our calculations with the tables containing a) data for the number density, flow velocity and temperature for the mixture with $Mach = 2, 3$ and concentrations $\chi_2^{(1)} = 0.1, 0.5, 0.9$; (b) grids in momentum and configuration spaces and integration nodes, together with
the analysis of their accuracy can be found in our paper \cite{Raines02}.

\begin{figure}
\begin{center}
\includegraphics[width=0.33\textwidth,height=!,angle=-90,origin=c]{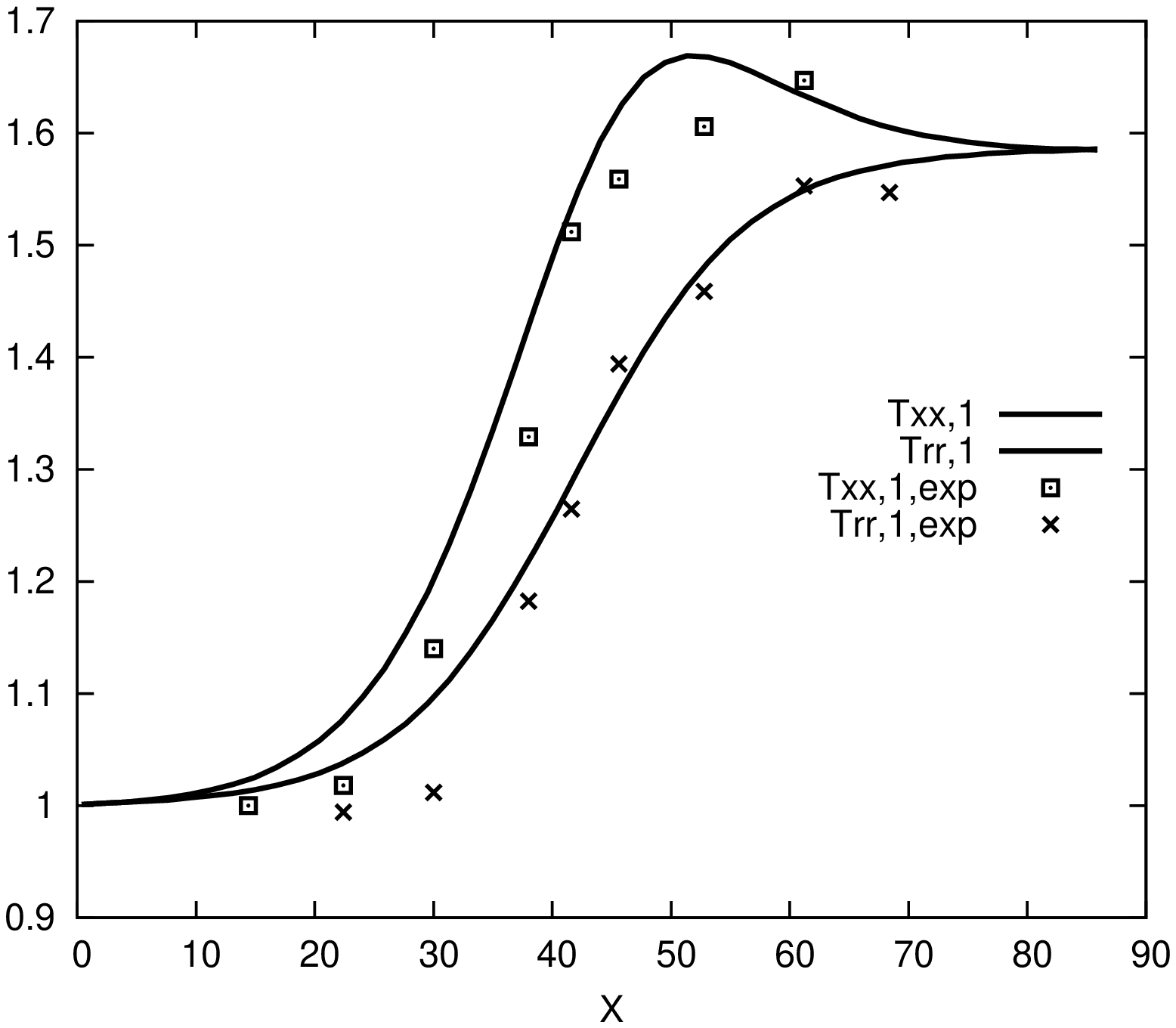}
\vspace*{-6mm}
\caption{\footnotesize Parallel and transversal temperatures for Argon (gas 1)\newline\hspace*{9mm} at $M = 1.58$,
$m_2/m_1 = 0.1$, $d_2/d_1 = 0.593$, $\chi_2^{(1)} = 0.9$.}
 \label{fig:9}
 \end{center}
\end{figure}

\begin{figure}
\begin{center}
\includegraphics[width=0.33\textwidth,height=!,angle=-90,origin=c]{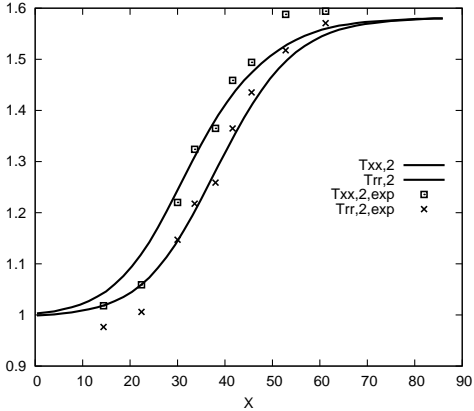}
\vspace*{-6mm}
\caption{\footnotesize Parallel and transversal temperatures for Helium\newline (gas 2) at $M = 1.58$,
$m_2/m_1 = 0.1$, $d_2/d_1 = 0.593$, $\chi_2^{(1)} = 0.9$.}
 \label{fig:10}
 \end{center}
\end{figure}

Fig. \ref{fig:9} and Fig. \ref{fig:10} show a comparison of our results for parallel $(T_{xx,i})$ and transversal $(T_{rr,i})$ temperatures for real gases Argon (gas 1) and Helium (gas 2) with experiments \cite{Harnett}. The results of \cite{Harnett} are marked by squares for $T_{xx,i}$ and by crosses for $T_{rr,i}$ while our results are denoted by solid lines. This shows a good agreement of our results with experimental data.

\begin{figure}
\begin{center}
\includegraphics[width=0.4\textwidth,height=!,angle=-90,origin=c]{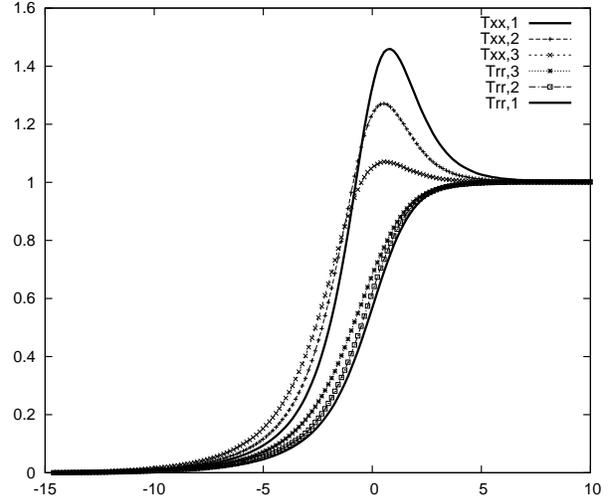}
\vspace*{-9mm}
\caption{\footnotesize Temperature tensor for Argon (gas 1), Nitrogen (gas 2),
\newline\hspace*{9mm} Methane (gas 3), Helium (gas 4)}
 \label{fig:11}
 \end{center}
\end{figure}

 \begin{figure}
 \begin{center}
\includegraphics[width=0.4\textwidth,height=!,angle=-90,origin=c]{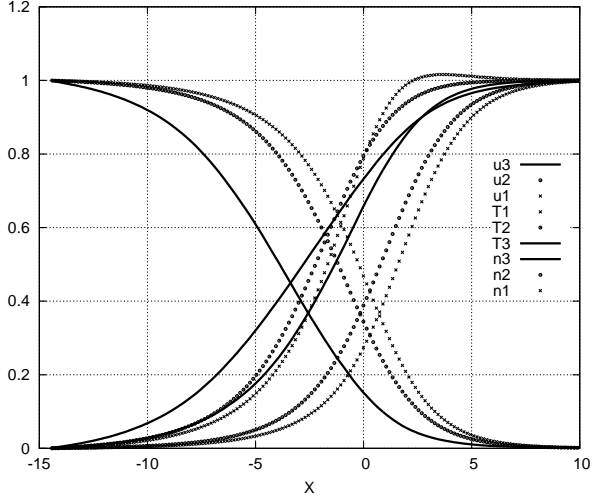}
\vspace*{-9mm}
\caption{\footnotesize Shock wave structure in the mixture of real gases: Argon\newline\hspace*{9mm} (gas 1),
Neon (gas 2), and Helium (gas 3).}
 \label{fig:12}
 \end{center}
\end{figure}

Fig. \ref{fig:11} and Fig. \ref{fig:12} present computational results for the three component mixture of monatomic gases: Argon (gas 1), Neon (gas 2) and Helium (gas 3), with real masses and molecular diameters. Computations were made for the following parameters of the shock wave components: $M=3, m_2/m_1=0.1, m_3/m_1=0.1$, $d_2/d_1=0.7, d_3/d_1=0.6, \chi_1^{(1)}=0.2, \chi_2^{(1)}=0.3, \chi_3^{(1)}=0.5$. On Fig. \ref{fig:11} one can see the maximum hump of the parallel temperature of Argon and, accordingly, on Fig. \ref{fig:12} the overshoot of the total temperature of Argon (the same  phenomenon can be seen in Fig. \ref{fig:5} and Fig. \ref{fig:6} for a binary gas mixture) together with a big difference between the graphs of all the mixture components.

In \cite{Jos} 3-component mixture has the following parameters: $M=3$, $m_1:m_2:m_3 = 1:0.9:0.8$, $d_1:d_2:d_3=1:1:1$, $\chi_1:\chi_2:\chi_1=1:2:3$. In figures presented in this paper one can see (in our notation) the coincidence of $n_i$ and $n$, $u_i$ and $u$, $T_{rr,i}$ for all $i=1,2,3$ and a little difference between parallel temperatures of components $T_{xx,i}$.

 \begin{figure}
  \begin{center}
\includegraphics[width=0.4\textwidth,height=!,angle=-90,origin=c]{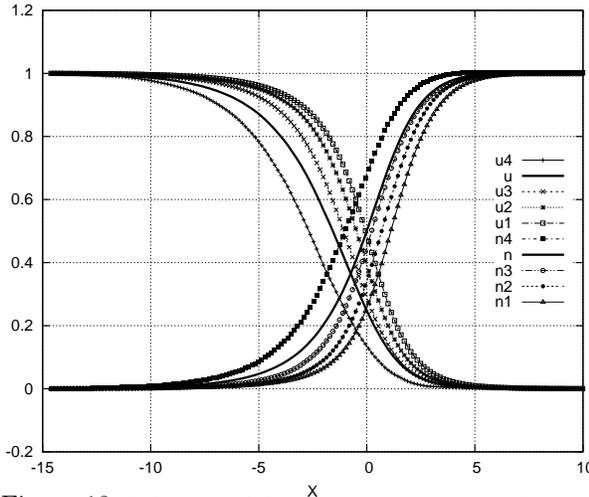}
\vspace*{-11mm}
\caption{\footnotesize Velocity and density for Argon (gas 1), Nitrogen (gas 2),
Methane (gas 3), Helium (gas 4) and for the mixture as a whole.}
 \label{fig:13}
  \end{center}
\end{figure}

 \begin{figure}
  \begin{center}
\includegraphics[width=0.4\textwidth,height=!,angle=-90,origin=c]{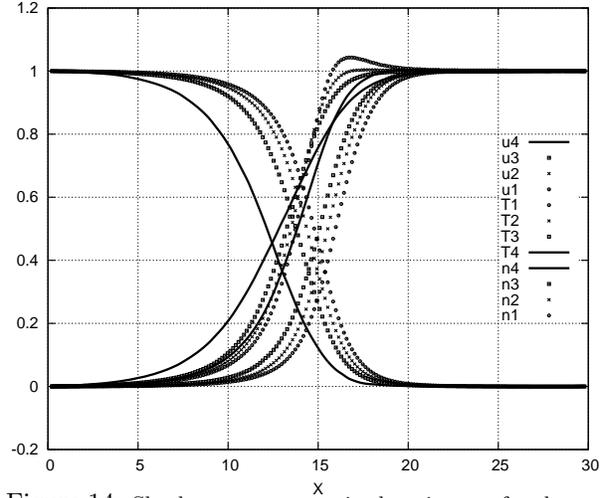}
\vspace*{-11mm}
\caption{\footnotesize Shock wave structure in the mixture of real gases: Argon (gas 1), Nitrogen (gas 2),
Methane (gas 3), and Helium (gas 4).}
 \label{fig:14}
  \end{center}
\end{figure}

Fig. \ref{fig:13} and Fig. \ref{fig:14} show the results of computations for 4-component mixture of Argon (gas 1), Nitrogen (gas 2),
Methane (gas 3) and Helium (gas 4). Molecular masses and daimeters of the components are taken real but the internal energies of Nitrogen and Methane are not taken into account. Computations are made for $M=3$ and the following parameters of the mixture:\\ $m_2/m_1=0.7$, $m_3/m_1=0.4$, $m_4/m_1=0.1$,\\ $d_2/d_1=1.034$, $d_3/d_1=1.144$, $d_4/d_1=0.6$,\\ $\chi_1^{(1)}=0.1$, $\chi_2^{(1)}=0.2$, $\chi_3^{(1)}=0.3$, $\chi_4^{(1)}=0.4$.\\
In Fig. \ref{fig:13} one can see that the graphs of the total density and velocity of the mixture lie between profiles of the components. In Fig. \ref{fig:14} one can see the overshoot of total temperature of Argon (gas 1) and a big difference between the graphs of all the mixture components.

 Since on Fig. \ref{fig:12} and Fig. \ref{fig:14} the ratios of molecular masses for the last component
differ too much $(m_i/m_1=0.1,\ i = 3, 4)$, we see that the profile of temperature of the last (lightest) component is flat. If the mass ratio of the last component differs not so much from the others, we do not see so flat profiles. All the calculations were performed on the laptop computer Sony VAIO, processor Intel(R) Core(TM)2CPU, 1.66GHz+1.66GHz, 1.00GB of RAM. For these calculations we have taken the following values of parameters: $20000$ nodes of the momentum grid with the step $h = 0.08$, $200$ nodes of the $x$-grid with the step $h_x = 0.15$, $66000$ integration nodes and $\Delta t = 0.01$.

Thus, we have made calculations for two, three and four components of the mixture. The test calculations were made for binary gas mixtures with the concentration $\chi_2^{(1)}$ of the second component: $0.1, 0.5, 0.9, 0.95$, ratio of masses $0.1, 0.25, 0.5$, Mach number $1.5, 2, 3, 6, 8$.

The method of \cite{Tcherem98} was extended for the gases with internal degrees of freedom in \cite{Tcherem05}, where it can incorporate real physical parameters of molecular potentials and internal energy spectrum.

\section{Conclusions}

Computational results for the shock wave structure presented in this paper were obtained by a unique approach based on the application of the Conservative Projection Method (CPM) \cite{Cherem97,Tcherem98,Tcherem00,Tcherem06} for solving the classical kinetic Boltzmann equation for monatomic gases. This method was extended to binary gas mixtures in cylindrical coordinates and, later on, to three and four component gas mixtures \cite{RainesEu,Raines02,Raines09} and in this paper. This method ensures strict conservation of mass, momentum and energy.
The transition from the upstream to downstream state was presented by distribution functions and their moments (macroscopic values) for various parameters (Mach numbers, ratios of masses and concentrations). Details of our calculations together with analysis of their accuracy can be found in paper \cite{Raines02}. With this
method we can obtain results on rough grids which coincide well with the results obtained on more fine grids. The numerical results have been compared with numerical and experimental results of other authors with a good agreement with them. All computations were performed on personal computers without using parallel processing. We have shown that the projection method for a gas mixture solves the shock wave problem with acceptable precision, small time of calculations and small computer memory.

\section*{Acknowledgements}

The author is grateful to prof. F.G. Tcheremissine for fruitful advices
and helpful remarks.


\begin{thebibliography}{99}
\bibitem{RainesEu}
Raines, A.A.: Study of a shock wave structure in gas mixtures on the basis of the Boltzmann equation. Eur. J. Mech. B Fluids \textbf{21}, 599--610 (2002)
\bibitem{nord}
Nordsieck, A., Hicks, B.L.: Monte-Carlo evaluation of the Boltzmann collision integral. In: Rarefied Gas Dynamics, Vol. \textbf{1}, Plenum Press, New-York-London, 695--710 (1967)
\bibitem{Tcherem70}
Tcheremissine, F.G.: Numerical solution of the Boltzmann kinetic equation for one-dimensional steady gas flows. J. Comp. Math. and Math. Phys. \textbf{10}, 654--665 (1970) (in Russian)
\bibitem{Oh93}
Ohvada, T.: Structure of normal shock waves. Direct numerical analysis of the Boltzmann equation for hard-sphere molecules. Phys. Fluids A \textbf{5}, 217--234 (1993))
\bibitem{Oh94}
Ohvada, T.: Numerical analysis of normal shock waves on the basis of the Boltzmann equation for hard-sphere molecules. In: Eds. Shizgal, B.D., Waver, D.P.: Rarefied Gas Dynamics: Theory and Simulations, IAA, Washington, p. 482 (1994)
\bibitem{Arist80}
Aristov, V.V., Tcheremissine, F.G.: The conservative splitting method for solving Boltzmann's equation. J. Comp. Math. and Math. Phys. \textbf{20}, 191--207 (1980) (in Russian)
\bibitem{Cherem97}
Cheremisin, F.G.: A conservative method of evaluation of a Boltzmann collision integral. Doklady RAN \textbf{357}, 53--56 (1997)
\bibitem{Tcherem98}
Tcheremissine, F.G.: Conservative evaluation of Boltzmann
collision integral in discrete ordinate approximation. Comp. Math. Appl. \textbf{35}, 215--221 (1998)
\bibitem{Tcherem00}
Tcheremissine, F.G.: Solution of the Boltzmann equation for transition to hydrodynamic regime. Doklady RAN \textbf{373}, 483--486 (2000)
\bibitem{Tcherem06}
Tcheremissine, F.G.: Solution of the Boltzmann kinetic equation for high speed flows. J. Comp. Math. and Math. Phys. \textbf{46}, 315--329 (2006)
\bibitem{Kosuge}
Kosuge, S., Aoki, K., Takata, S.: Shock-wave structure for a binary gas mixture: finite-difference analysis of the Boltzmann
equation for hard sphere molecules. Eur. J. Mech. B Fluids \textbf{20}, 87--126 (2001)
\bibitem{Maus}
Mausbach, P., Beylich, A.E.: Numerical solution of the Boltzmann equation for one-dimensional problems in binary mixtures. In: Proc. 13 Internat. Symp. Rarefied Gas Dynamics, Vol. \textbf{1}, Plenum Press, New York, 285--293 (1985)
\bibitem{Raines91}
Raines, A.A.: Numerical solution of the Boltzmann equation for one-dimensional problem in a binary gas mixture. In: Ed. Beylich, A.E.: Proc. 17 Internat. RGD Symp., VCH, Weiheim-New-York, 328--331 (1991)
\bibitem{Raines02}
Raines, A.A.: A method for solving the Boltzmann equation for a gas mixture in the case of cylindrical symmetry in the velocity space. J. Comp. Math. and Math. Phys. \textbf{42}, 1212--1223 (2002)
\bibitem{Raines09}
Raines, A.A.: Numerical solution of the Boltzmann equation for the shock wave in a gas mixture. In: 27 Internat. Symp. on Shock Waves, Book of Proceedings, St.Petersburg, p. 213 (2009)
\bibitem{Jos}
Josyula, E., Vedula, P., Bailey, W.F.: Kinetic solution of shock structure in a non-reactive gas mixture. In: 48th AIAA Aerospace Sciences Meeting, AIAA 2010-817, Orlando, Florida (2010)
\bibitem{korob}
Korobov, M.M.: Trigonometric Sums and Their Applications, pp. 1--240. Mir, Moscow (1989)
\bibitem{Boris}
Boris, J.P., Book, D.L.: Flux-corrected transport. 1. SHASTA, A fluid transport algorithm that works. J. Comp. Phys. \textbf{11}, 38--69 (1979)
\bibitem{Harnett}
Harnett, L.M., Muntz, E.P.: Experimental investigation of normal shock wave velocity distribution functions in mixtures of Argon
and Helium. Phys. Fluids \textbf{15}, 565--572 (1972)
\bibitem{Bey}
Beylich, A.E.: Kinetic model for the shock structure in a binary gas mixture. Phys. Fluids \textbf{11}, 2764 (1968)
\bibitem{bird68}
Bird, G.A.: The structure of normal shock waves in a binary gas mixture. J. Fluid Mech. \textbf{31}, 657 (1968)
\bibitem{bird94}
Bird, G.A.: Molecular Gas Dynamics and Direct Simulation of Gas Flows, Oxford Univ. Press, Oxford (1994)
\bibitem{bird84}
Bird, G.A.: Shock wave structure in gas mixtures. In: Ed. Oguchi,
H.: Rarefied Gas Dynamics, Vol. \textbf{1}, Univ. Tokyo Press,
Tokyo, pp. 175--184 (1984)
\bibitem{Tcherem05}
Tcheremissine, F.: Direct numerical solution of the Boltzmann
equation. In: 24th Intern. Symp. Rarefied Gas Dynamics, AIP
Conference Proceedings, pp. 667--685 (2005)
\end{thebibliography}
\end{document}